\documentclass{emulateapj}
\usepackage{graphicx}
\usepackage{amsmath}
\begin{document}

\def\etal{et al.\ \rm}
\def\ba{\begin{eqnarray}}
\def\ea{\end{eqnarray}}
\def\etal{et al.\ \rm}

\title{Analysis of Spin-Orbit Misalignment in Eclipsing Binary DI Herculis}

\author{Alexander A. Philippov\altaffilmark{1,2} \& Roman R. Rafikov\altaffilmark{2,3}}
\altaffiltext{1}{Moscow Institute of Physics and Technology, Institutsky per., 9, Dolgoprudny, Moscow region, 141700, Russia}
\altaffiltext{2}{Department of Astrophysical Sciences, 
Princeton University, Ivy Lane, Princeton, NJ 08540; 
rrr@astro.princeton.edu}
\altaffiltext{3}{Sloan Fellow}


\begin{abstract}
Eclipsing binary DI Herculis (DI Her) is known to exhibit 
anomalously slow apsidal precession, below the rate
predicted by the general relativity. Recent measurements 
of the Rossiter-McLauglin effect indicate that
stellar spins in DI Her are almost orthogonal to the 
orbital angular momentum, which explains the anomalous 
precession in agreement with the earlier theoretical suggestion 
by Shakura. However, these measurements yield only the 
projections of the spin-orbit angles onto the 
sky plane, leaving the spin projection onto our line of sight 
unconstrained. Here we describe a method of determining 
the full three-dimensional spin orientation of the binary
components relying on the use of the gravity darkening 
effect, which is significant for the rapidly rotating 
stars in DI Her. Gravity darkening gives rise to nonuniform 
brightness distribution over the stellar surface, the pattern 
of which depends on the stellar spin orientation. Using 
archival photometric data obtained during multiple 
eclipses spread over several decades we are able to constrain 
the unknown spin angles in DI Her with this method, finding 
that spin axes of both stars lie close to the plane of the 
sky. Our procedure fully accounts for the precession of 
stellar spins over the long time span of observations.  
\end{abstract}



\section{Introduction.}  
\label{sect:intro}


The binary system DI Herculis (DI Her, HD 175227) was discovered 
as an eclipsing variable by Hoffmeister (1930). It consists of 
two massive B stars on an eccentric orbit ($e = 0.49$) with 
period $P = 10^{d}.55$, inclined at an angle $i = 89.3^{\circ}$
with respect to our line of sight (see Table \ref{tbl:pars}
for parameters of both binary components). A unique property of this
system that has been attracting a lot of attention for almost 
three decades is its low rate of apsidal precession, 
$\dot\omega_{obs}=1.24^\circ \pm 0.18^\circ /100$ yr 
(Martynov \& Khaliullin 1980). This is almost two times 
lower than the general relativistic apsidal precession rate 
$\dot\omega_{\rm GR}=2.35^\circ/100$ yr theoretically predicted
based on the measured parameters of the system
(Rudkjobing 1959). For a long time 
this discrepancy was not understood (Maloney \etal 1989) 
and even ascribed to the failure of general relativity
(Moffat 1989).

Shakura (1985) suggested that slow apsidal precession 
in DI Her is caused by the misalignment between the spin 
and orbital angular momentum axes of the system. Indeed, 
for stellar spins strongly misaligned with the orbital angular 
momentum the rotation-induced stellar quadrupole gives 
rise to a contribution to $\dot \omega$ with a sign opposite 
to that of $\dot\omega_{\rm GR}$. Then, if this 
quadrupole-induced precession 
is fast enough it can easily alter the full rate of apsidal 
precession and even make it smaller than $\dot\omega_{\rm GR}$
as in the case of DI Her. Somewhat less extreme version of this
idea has recently been applied to another eclipsing binary 
AS Camelopardalis (Pavlovski \etal 2011), which also exhibits
relatively slow rate of apsidal precession. 

The spin-orbit misalignment in DI Her has been recently confirmed 
by Albrecht \etal (2009) who used the evolution of stellar spectral 
signatures during the eclipse (the so-called Rossiter-McLaughlin 
effect, Holt 1893; Rossiter 1924; McLaughlin 1924) to set 
constraints on the spin orientation of both stars. They 
found that both stars of DI Her have 
their spin axes nearly perpendicular to the orbital angular 
momentum axis, which is at odds with the common wisdom 
regarding spin-orbit orientation in close binary stars, but 
can naturally explain the slow apsidal precession in this system. 

Unfortunately, the Rossiter-McLaughlin effect allows one to
determine only the {\it sky plane projection} $\lambda$ of 
the angle $\alpha$ between the spin and orbital momentum 
axes for the stars. The angle $\beta$ between the stellar spin 
and our line of sight remains essentially unconstrained. However, 
figuring out whether spin-orbit misalignment can explain the 
observed $\dot \omega$ does depend on the value of $\beta$,
since the stellar spin frequency $\omega$ is inferred from the 
measured projected stellar rotation speed 
$v_{rot}\sin\beta=\omega R_\star\sin\beta$ ($R_\star$ 
is the stellar radius). Albrecht \etal (2009) and Claret 
\etal (2010) tackled the issue of undetermined $\beta$ by 
means of Monte Carlo simulations, assuming this angle 
to be uniformly distributed. 

In this work we develop a method of analyzing the photometric
eclipse data, which allows us to constrain the angle $\beta$
without using spectroscopic data. This method relies on the fact 
that both components of DI Her are rapidly rotating stars 
($v_{rot}\sin\beta$ exceeds $100$ km s$^{-1}$ for both 
components, see Table \ref{tbl:pars}),
and must exhibit a non-uniform surface brightness distribution 
due to the gravity darkening effect (von Zeipel 1924). 
This surface brightness pattern is sensitive to the orientation 
of stellar spin axis with respect to our line of sight. By 
probing the brightness distribution via the detailed shape 
of the system lightcurve during the eclipse one can infer the
full spin orientation of both stars. Analogous 
method was recently proposed by Barnes (2009) for 
analyzing planetary transits around rapidly rotating stars,
and applied by Szab\'o \etal (2011) and Barnes \etal 
(2011) to determine the spin-orbit misalignment in a transiting 
system KOI-13.01.

In determining the unknown angles $\beta$ for both stellar 
components we also use other constraints on the system 
parameters, such as the observed apsidal precession rate 
and the evolution of the projected stellar rotation 
velocities $v_{rot}\sin\beta$ over the long time span. 

This work is organized as follows. In \S \ref{sect:lc} we 
describe geometric setup of the problem, and eclipse 
lightcurve modeling. In \S \ref{sect:obs_fits} we 
describe observational data and our fitting procedure. Our 
results are presented in \S \ref{sect:results} and discussed 
in \S \ref{sect:discussion}. In Appendix \ref{app:evolve}
we derive equations describing evolution of the system 
orientation as a result of spin and orbital precession, 
which may find other applications.


\section{Eclipse modeling.}
\label{sect:lc}



\subsection{Geometry of the system.}
\label{sect:geom}

\begin{figure}
\begin{center}
\epsscale{1.28}
\plotone{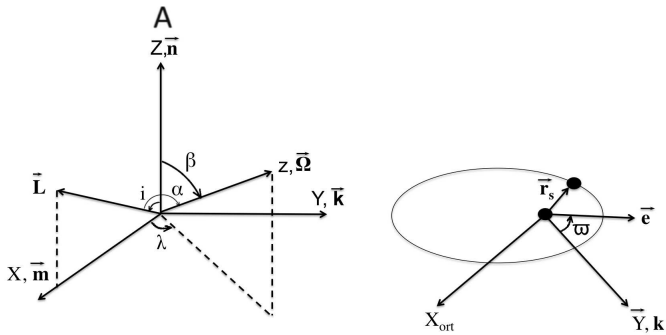}
\caption{Observer coordinate frame on the left, and a sketch of
the orbital geometry on the right. See text for 
more details.}
\label{fig1}
\end{center}
\end{figure}

To model eclipse lightcurve we use two Cartesian coordinate 
systems. One is the {\it observer} 
frame $(X,Y,Z)$, which describes the orbital orientation of 
the system:  $Z$ axis points from the system barycenter 
towards the observer, $X$ axis is along the direction of 
the sky-projected orbital angular momentum and $Y$ axis is 
aligned with the line of nodes, see Figure \ref{fig1}.
 
Another system $(x_i,y_i,z_i)$, where $i=p,s$ for primary and 
secondary, respectively, is aligned with the stellar symmetry 
axis ({\it symmetry} frame): its $z_i$ axis is along the stellar 
spin angular velocity ${\bf \omega}_j$, and $x_i$ and $y_i$ axes 
are obtained from $X$ and $Y$ by two rotations: first, a rotation 
around $Z$ axis by the angle $\lambda_{i}$ and then another 
rotation around axis obtained from $Y$ in previous step by the angle 
$\beta_{i}$. We will use the symmetry frame to describe 
the stellar surface shape and temperature distribution and 
the observer frame to characterize the visible sky-projected 
stellar disc.

Spin angular velocity $\boldsymbol{\omega}_j$ in the observer 
frame is then given by
\begin{equation}
{\boldsymbol{\omega}}_{j}=
\begin{pmatrix} 
\sin \beta_{j} \cos\lambda_{j} \\ 
\sin\beta_{j}\sin\lambda_{j}\\
\cos\beta_{j} \\
\end{pmatrix}
\end{equation}

The vector towards the observer in the symmetry system 
of each star is
\begin{equation}
\boldsymbol{n}=
\begin{pmatrix}
-\sin \beta_{i} \\ 
0 \\ 
\cos \beta_{i} \\
\end{pmatrix}
\label{observ}
\end{equation}

To simulate the eclipse light curves we also need to know the relative 
stellar trajectory projected onto the plane of the sky. 
The projected position of the center of the secondary with respect
to primary in the observer frame is given by
\begin{equation}
{\bf R_s}=r_s
\begin{pmatrix}
-\cos i\sin(f+\varpi)\\ 
\cos(f+\varpi)\\ 
\sin i\sin(f+\varpi)
\end{pmatrix}
\label{eq:Rs}
\end{equation}
where $r_{s}$ is the distance of the secondary star with respect to the primary and $f$ is the true anomaly (see Murray \& Dermott 2000 for relation with other orbital parameters).
Equation (\ref{eq:Rs}) fully determine the 
time evolution of ${\bf R_s}$ during the eclipse.


\subsection{Shape of the stellar surface.}
\label{sect:shape}

We now describe the variation of intensity of emission over 
the stellar surface due to the gravity darkening effect.
Our results will be valid both for the primary and the secondary,
so we will omit the subscript $i=p,s$ in this subsection.

First, we determine the geometry of the sky-projected disc by 
assuming the shape of the stellar surface to coincide with 
isopotential surfaces for the effective potential
\begin{equation}
\Phi_{\rm eff}(x,y,z) = -\frac{GM_\star}{\sqrt{x^2+y^2+z^2}} 
- \frac{1}{2} \Omega_\star^2 (x^2 + y^2),
\end{equation}
as a function of coordinates in the "symmetry" frame. Assuming slow 
rotation one can obtain the equation for the stellar surface in
the form
\begin{equation}
\frac{x^2+y^2}{{\eta}^2}+z^2={R^2_{pol}},
\label{form}
\end{equation}
where $R_{pol}$ is the polar radius of the star. 
Thus, because of the rotation stellar surface has an 
ellipsoidal shape with oblateness 
$\eta$ given by
\begin{equation}\eta=\frac{R_{eq}}{R_{pol}} = 
1 + \frac{R^3_i\omega^2_i}{2GM_i} = 1 + \frac{S}{2}.
\end{equation}
Here the parameter S is related to the ratio of the rotation 
rate $\omega_{i}$ to the breakup rotation rate 
$\omega_{b}\equiv (GM_\star/R_\star^3)^{1/2}$ at which the 
centrifugal force balances gravity at the stellar surface.
Equation (\ref{form}) can be re-written in polar coordinates 
as $r(\theta) = R_{pole} (1 + S \sin^2\theta/2)$.

The actual angular velocity of stellar spin is calculated from 
the spectroscopically measured projected stellar rotation velocity 
$(v_{rot} \sin\beta)_{obs}$ as
\begin{equation}
\omega = \frac{(v_{rot} \sin\beta)_{obs}}{R_{eq} \sin\beta},
\label{eq:vsin}
\end{equation}
where we assumed $R_{eq}$ equal to the stellar radius quoted
in the literature.


\subsection{Temperature distribution over the stellar surface.}
\label{sect:Tdist}

Because of rapid stellar rotation the brightness temperature 
of the stellar surface is not constant but obeys the 
von Zeipel (1924) law
\begin{equation}
T({\bf R})=T_{pol}\left[\frac{g_{\rm eff}({\bf R})}
{g_{pol}}\right]^{\beta_g}=
T_{pol}\psi({\bf R}),
\label{eq:vonZeipel}
\end{equation}
where $T_{pol}$ is the value of $T$ at the stellar pole, 
$\beta_g$ is the gravity darkening power law index, 
${\bf R}$ is the three-dimensional radius-vector from the 
stellar center to a point on the stellar surface, and 
$g_{\rm eff}$ is the local effective gravitational acceleration:
\begin{equation}
{\bf g}_{\rm eff}({\bf R})=-\frac{GM_{\star}}{R^3}{\bf R}+
\omega^2 {\bf R_{\perp}}.
\label{eq:geff}
\end{equation}
Here ${\bf R}_{\perp}={\bf R} - (\boldsymbol{\omega}\cdot{\bf R})/\omega$ 
is the distance to that point from the stellar spin axis, and 
$g_{pol}$ is the value of $g_{\rm eff}$ at the stellar pole, where 
$R_{\perp}=0$. 

\begin{figure}
\epsscale{1.0}
\plotone{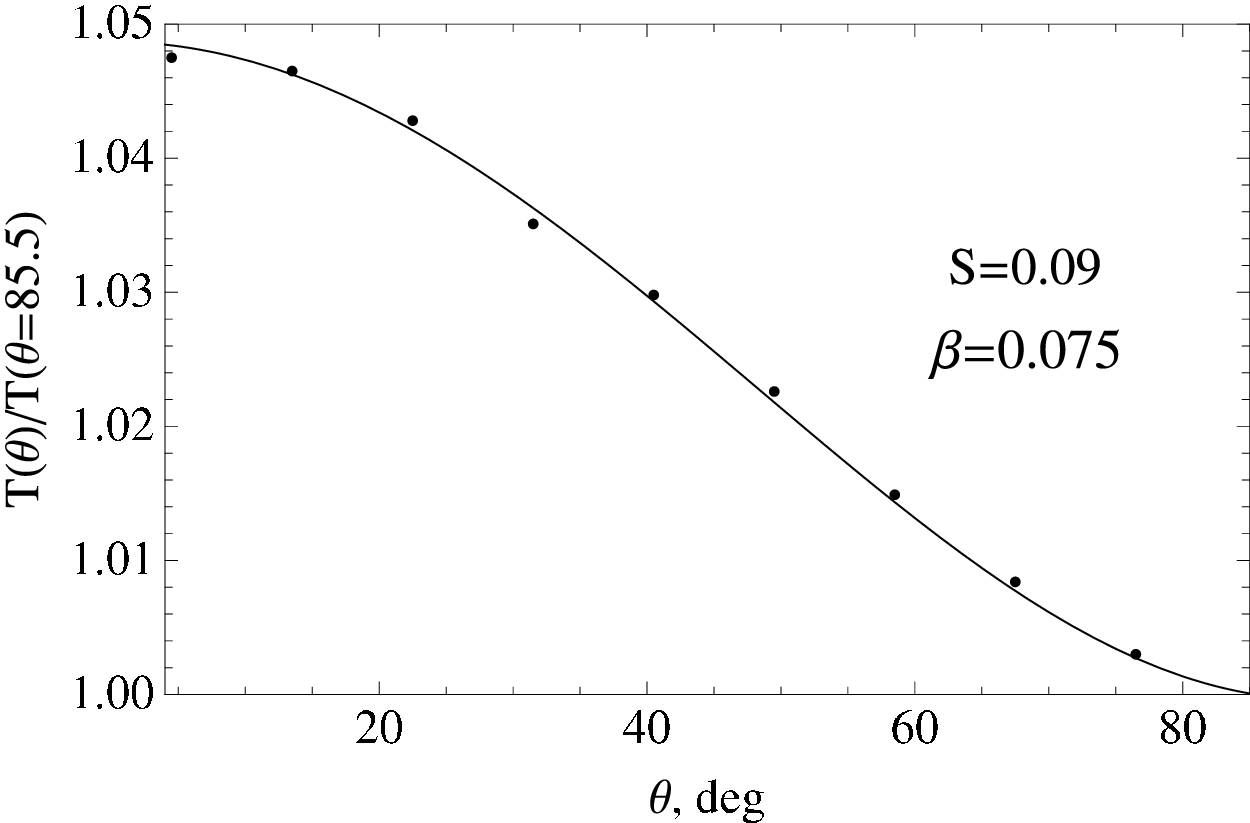}
\caption{Latitudinal distribution of the brightness temperature 
for a stellar model with rotation parameter $S = 0.09$ (points; 
Deupree, 2010) fitted by a von Zeipel law (equation 
(\ref{eq:vonZeipel}); solid line) with low value of 
$\beta_g = 0.075$ that is close to $\beta_g = 0.1$  
used in this work.}
\label{fig12}
\end{figure}

Conventional gravity darkening theory (von Zeipel 1924) predicts 
$\beta_g=0.25$. However, recent detailed 
theoretical calculations (Deupree 2011) of the latitudinal distribution 
of the effective temperature for rotating stars performed in the wide 
range of stellar masses (between 1.625 and 8 $M_{\odot}$) suggest a 
considerably weaker dependence of $T({\bf R})$ on $g_{\rm eff}$. We 
illustrate this point in Figure \ref{fig12}, where we plot the 
latitudinal distribution of the effective temperature for a 
particular stellar model from Deupree (2011) corresponding to a 
rotation parameter $S=0.09$. This distribution depends on stellar 
mass only weakly, meaning that it can be applied for DI Her 
components as well. One can see that equation (\ref{eq:vonZeipel}) 
with $\beta_g=0.075$, which is considerably lower than $0.25$, 
provides excellent fit to these data. 

On the observational side, interferometric measurements for 
rapidly rotating stars by Che \etal (2011) find $\beta_g\approx 0.146$ 
for 1.77$M_{\odot}$ star $\beta$ Cassiopeiae, having 
$v_{rot}\sin\beta\approx 75$ km s$^{-1}$ and $\beta_g\approx 0.19$ 
for 4.15$M_{\odot}$ star $\alpha$ Leo, rotating at 
$v_{rot}\sin\beta\approx 340$ km s$^{-1}$. Even though the latter 
is very similar in mass to the DI Her components, it spins much 
faster (spin parameter $S$ is almost an order of magnitude higher), 
making direct extrapolation to the DI Her case difficult. Despite
these ambiguities, it is clear that both theoretically and 
observationally one typically infers $\beta_d<0.25$. In this work 
we have chosen to adopt $\beta_g=0.1$ more in line with the work of 
Deupree (2011).

\subsection{Intensity distribution over the sky-projected disc.}
\label{sect:int_dist}

\begin{figure}
\epsscale{1.25}
\plotone{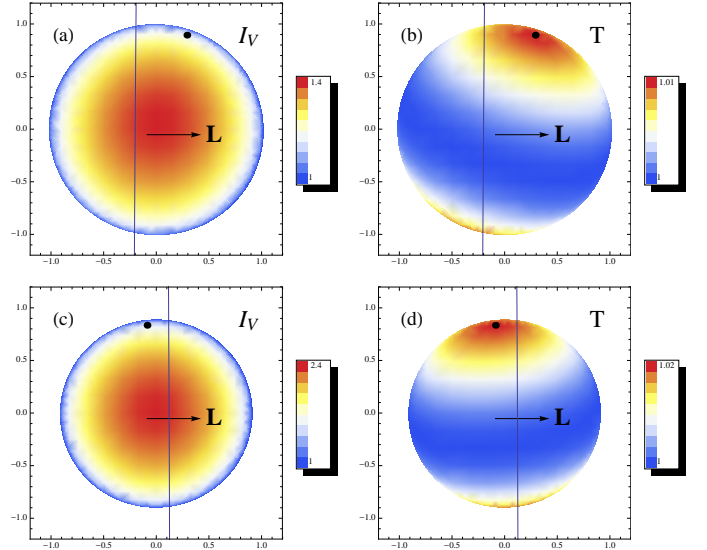}
\caption{Intensity distribution for the primary (a) 
and secondary (c) stars, and temperature distribution for the 
primary (b) and secondary (d). Calculations assume $\beta_p=70^{\circ}$, 
$\lambda_p=72^{\circ}$, $\beta_s=110^{\circ}$, 
$\lambda_s=-84^{\circ}$, $S=0.040$ for the primary 
star, $S=0.046$ for the secondary star, limb-darkening coefficient $c_1=0.35$ 
for the primary star, $c_1=0.64$ for the secondary star and 
von Zeipel parameter $\beta_g=0.1$. The dot on each plot 
indicates the position of the stellar pole. The blue line 
shows the trajectory of the center 
of the secondary (primary) star during primary (secondary) 
eclipse. The arrow shows the direction of the projected 
orbital angular momentum ${\bf L}$. Temperature distribution 
clearly illustrates gravity darkening. The intensity variation 
is large (tens of per cent) mainly because of the limb 
darkening effect, which 
is much larger than the gravity darkening.}
\label{fig9}
\end{figure}

Equations (\ref{eq:vonZeipel}), (\ref{eq:geff}) provide us with 
a simple expression for $T({\bf R})$ in the symmetry frame of 
a star, since ${\bf R}$ in this frame can be trivially derived 
from equation (\ref{form}). However, for the purposes of 
eclipse lightcurve modeling we need to know the temperature 
distribution in the {\it observer} frame,
projected onto the plane of the sky. In Appendix \ref{eq:conv} 
we describe the relation between $(x_i,y_i,z_i)$ and $(X,Y,Z)$
frames, which allows us to write the dimensionless function $\psi$
in equation (\ref{eq:vonZeipel}) as 
$\psi({\bf R})=\psi(X,Y,Z(X,Y,\boldsymbol{\omega}_i))=
\psi(X,Y,{\bf \omega}_i)$ for each star.
The dependence of $\psi$ on $\boldsymbol{\omega}_i$ is the key factor that 
allows us to use stellar photometry during eclipse to determine the
spin orientation of the stars. Figure \ref{fig9}b,d illustrates
the distribution of the brightness temperature over the stellar
surface projected onto the sky plane. 

Spectral density of the stellar radiation flux detected on 
Earth is
\begin{equation}
F_{\lambda} = \frac{1}{d^2} \int\limits_{S} 
I_{\lambda}(T_{pol} \psi (X,Y,\boldsymbol{\omega})) 
\Phi_{\lambda}(X,Y,\boldsymbol{\omega}){dX dY},
\label{flux}
\end{equation}
where $d$ is the distance to the system, $\Phi_{\lambda}$ is
the limb-darkening law, and $I_{\lambda}(T)$ is the spectral 
intensity at a given temperature $T$. In this work we assume 
that $I_{\lambda}(T) = B_{\lambda}(T)$, where
$B_{\lambda}(T)$ is a standard black-body radiation function. 
Thus, spectrum of 
each star is in general a multi-color blackbody, parametrized 
by $T_{pol}$ and $\psi$.

On the other hand, values of the effective temperature $T_\star$
for both stars quoted in the literature are obtained assuming that
both stars radiate as pure blackbodies characterized by a single
value of temperature, uniform across the stellar surface 
(Claret 2010). In that case the total 
bolometric flux is $F = L_{*}/(4\pi d^2) 
= \sigma T^4_{eff} R^2_{*}/d^2$, where $L_{*}$ is the 
stellar luminosity. This assumption is not valid for rapidly
rotating stars because of gravity darkening, and $T_{pol}$
cannot be taken equal to $T_\star$. To relate them
we integrate equation (\ref{flux}) over all wavelengths to 
obtain the following expression for $T_{pol}$:
\begin{equation}
T_{pol}=T_\star \left[\frac{1}{\pi R^2_{*}}\int\limits_{S} 
\psi^4 (X,Y,\boldsymbol{\omega}) \Phi_{\lambda}(X,Y,\boldsymbol{\omega})
{dX dY}\right]^{-1/4},
\end{equation} 
i.e. there is a correction depending on the orientation of stellar 
spin. This self-consistent derivation of $T_{pol}$ is an important 
part of our procedure which distinguishes it from the
approach of Barnes (2009).

For simplicity the limb-darkening law in this work is 
assumed to be frequency- and spin-independent and have
a simple functional form
\begin{equation}
\Phi_\lambda = 1-c_{1}(1-\mu),
\label{intens1}
\end{equation}
where $c_1$ is constant and $\mu$ is the cosine of the 
angle between the local normal ${\boldsymbol {\xi}}$ to the 
stellar surface and the observer's line of sight:
$\mu = {\boldsymbol {\xi}}\cdot {\bf {n}}$.
The total measured flux in the V band $F_V$ is obtained by 
additionally convolving $F_\lambda$ in equation (\ref{flux}) 
with $W_{\lambda}$ --- the normalized transmission function 
for that band --- over $\lambda$. 

Figure \ref{fig9}a,c shows how the radiation 
intensity is distributed over the stellar surface for the 
two components of the DI Her system out of eclipse, when both 
gravity-darkening and limb-darkening are taken into account.
It is obvious that limb-darkening has a much stronger effect
on the intensity distribution than the gravity darkening, 
complicating measurement of the latter effect in the photometric 
data and determination of the spin orientation of the two stars.
On the other hand, as long as the stellar spin axis is not aligned 
with our line of sight, the gravity darkening results in 
a {\it non-axisymmetric} brightness distribution with respect 
to our line of sight, unlike the limb darkening, which is 
axisymmetric. This helps one disentangle the two contributions
in the photometric data. 

\begin{figure}
\epsscale{1.3}
\plotone{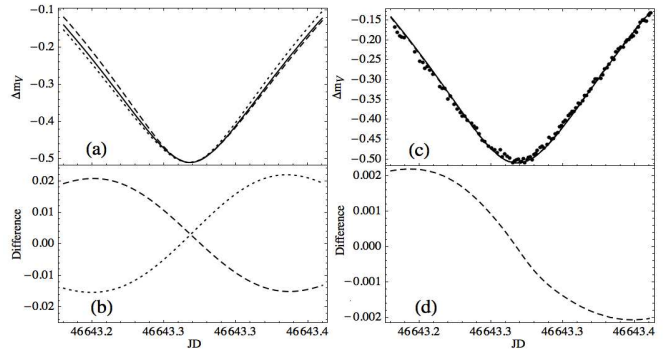}
\caption{(Left) Top panel shows simulated light curves of the secondary eclipse for the uniform temperature distribution (solid line), fast rotation (S=0.13) and two different secondary spin orientations - $\beta_{s1} = 70^{\circ}$, $\lambda_{s1} = 90^{\circ}$ (dashed line) and $\beta_{s2} = 70^{\circ}$, $\lambda_{s2} = -100^{\circ}$ (dotted line) (primary star parameters are kept constant at $\beta_{p} = 70^{\circ}$, $\lambda_{p} = 72^{\circ}$).  On the lower left panel the difference between simulated curves (dashed for the first case and dotted for the second one) and the case of uniform temperature distribution is given. For the first orientation dimmer regions are blocked first, with the opposite being true for the second one. (Right) Top panel shows simulated light curves for the uniform temperature distribution (dashed line) and actual DI Hercules parameters ($\beta_{p} = 70^{\circ}$, $\beta_{s} = 110^{\circ}$, $\lambda_{p} = 72^{\circ}$, $\lambda_{s} = -84^{\circ}$, $S_{p} = 0.040$, $S_{s} = 0.046$). Dots correspond to observational data (secondary eclipse 7/13/1986). The differences between the theoretical lightcurves are hardly visible, so we visualize them on the lower panel (we do not show data point there as they would be off scale).}
\label{fig11}
\end{figure}

For simulating the light curves we need to calculate the 
flux {\it blocked} during the eclipse
\begin{equation}
I_{bl,i}(\boldsymbol{\omega},t)=\int H (X,Y,t) F_{V,i}(X,Y,\boldsymbol{\omega})
dX dY,~~~i=p,s,
\label{eq:blocked}
\end{equation}
where $H(X,Y,t)$ equals 1, if the secondary (primary) star 
blocks starlight of the primary (secondary) at position $(X,Y)$, 
and 0 if not. Then the total flux observed on Earth is 
\begin{equation}
I(\boldsymbol{\omega},t)=I_{p} + I_{s} - I_{bl,i}(\boldsymbol{\omega},t),
\label{eq:intensity}
\end{equation}
where $I_p$ and $I_s$ are the unblocked stellar fluxes, i.e. calculated
from equation (\ref{eq:blocked}) with $H$ set to unity. 
During the eclipse $I_{bl,i}(\boldsymbol{\omega},t)$ 
changes because function $H$
varies with time across the surface of eclipsed star. In this work 
we take the surface (and frequency because of the finite bandwidth) 
integral in equation (\ref{eq:blocked}) using Monte Carlo 
technique. 

Our lightcurve modeling procedure is illustrated in Figure \ref{fig11},
where in left panels we demonstrate the differences between the
model assuming uniform distribution of the surface temperature and
the models in which the effect of gravity darkening is fully taken 
into account. One can see that for relatively fast rotation 
($S=0.13$ or $\omega=0.36\omega_b$) the difference between 
the uniform $T$ case and the gravity-darkened models is at the 
level of several per cent, which should be easily detectable in
single-epoch observations. 

In right panels of the same Figure we show the comparison between 
the uniform temperature case, the gravity-darkened model with spin
angles resulting from our fits to the data (see \S \ref{sect:results}).
In this case $S\approx 0.04$, see Table \ref{tbl:pars}, and the difference between 
the uniform and gravity-darkened models is very small, $\sim 10^{-3}$
mag. Thus, in the case of DI Her one would need very high-quality
photometry (at the level of several $\times 10^{-4}$ mag) to detect
gravity darkening-related asymmetries in the lightcurve shape.


\section{Observational data and fitting procedure}  
\label{sect:obs_fits}


\subsection{Observations.}  
\label{sect:obs}

The dataset we use for determining spin orientation of the 
DI Her components consists of photoelectric V band measurements 
of this system using the 50-cm AZT-14 reflector at the Tien 
Shan Observatory of the Astrophysical Institute of Kazakhstan 
and the Zeiss-600 reflector at the Crimean Station of the 
Sternberg Astronomical Institute prepared in 2003–2008. The 
database also contains the photoelectric observations 
going further in the past by Semeniuk (1968), the 1968–-1978 
observations by Martynov and Khaliullin (1980), the 1986–-1988 
observations by Khodykin, Volkov, and Metlov, and the 2004 
observations by Shugarov (see Kozyreva \& Bagaev 2009 and the 
references there in). These observations use different instruments, 
which were not cross-calibrated. Our fitting uses 9 eclipses, 
4 primary and 5 secondary. The photometric errors 
are unconstrained in all cases and we describe in 
\S \ref{sec:fit} how we deal with this issue.

\begin{figure*}
\epsscale{1.0}
\plotone{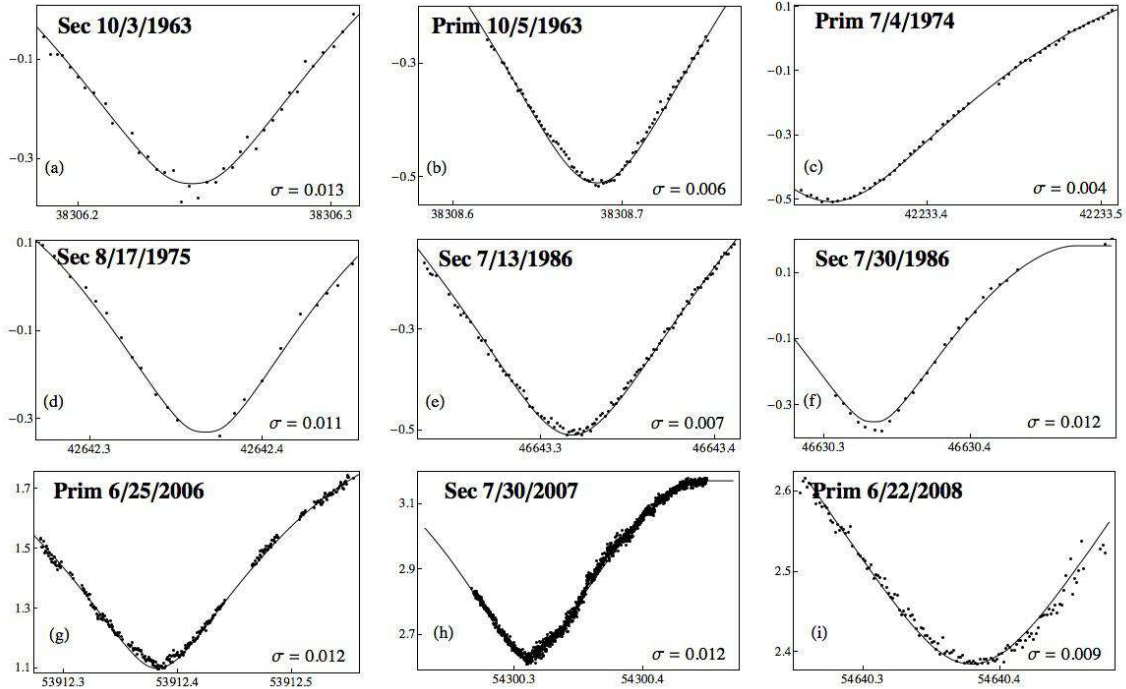}
\caption{Eclipse data used in our modeling with best fits 
overplotted. Horizontal axis is JD and the vertical one 
is apparent magnitude in the V band for every eclipse. 
Labels indicate the date of eclipses and the adopted noise 
levels.}
\label{fig5}
\end{figure*}


\subsection{Time evolution of the system.}  
\label{sect:time_evol}

\begin{figure}
\epsscale{1.2}
\plotone{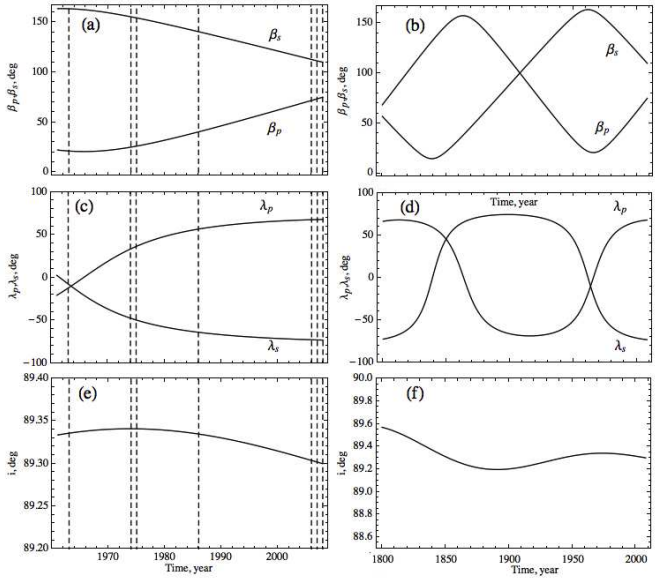}
\caption{
Evolution of the DI Her spin orientation (angles $\beta$ and 
$\lambda$) for both stars and the inclination $i$ of the system 
both on short (left) and long (right) time intervals. Dashed 
vertical lines correspond mark the locations of eclipses used 
in our modeling. Variation of spin angles $\beta_i$ and 
$\lambda_i$, $i=p,s$ are very significant.}
\label{fig0}
\end{figure}

High mass, rapid rotation and relative proximity of the 
stars in DI Her system drive rapid evolution of the spin
orientations for both components. In Appendix \ref{app:evolve}
we derive equations that describe evolution of the stellar
spins and orbital elements of the system. In particular, we 
show there that stellar spins in DI Her can rotate by more
than $100^\circ$ within a century, see equation 
(\ref{eq:Omega_P}). This evolution obtained by integrating 
equations from Appendix \ref{app:evolve} over time is 
illustrated in Figure \ref{fig0}.

This Figure clearly demonstrates that spin orientation
of the DI Her components significantly changes over the time
span of our full dataset. On one hand this complicates the eclipse
fitting, but on the other hand it provides us with a unique 
opportunity to use the signatures of this variation of stellar 
spins in eclipse modeling on long time intervals. 

In this regard our procedure of eclipse simulation uses somewhat 
different strategy that the one proposed by Barnes (2009): instead 
of using high-accuracy photometric data in a single epoch, which 
provides sensitivity to spin orientation only through the shape 
of the eclipse lightcurve, we use low-accuracy photometry obtained 
at different epochs, which allows for an additional effect of 
the current spin orientation of DI Her on the fitting procedure 
--- through the time evolution of ${\bf \omega}_i$.


\subsection{Evolution of $v_{rot}\sin\beta$.}  
\label{sect:vsini_evol}

Precession of stellar spins causes evolution of 
$v_{rot,i} \sin \beta_{i}$ ($i=p,s$) on long time interval, which
can be compared with past measurements. We took the values of 
$v_{rot,i} \sin \beta_{i}$ from Albrecht \etal (2009), who compiled 
measurements from different epochs starting in 1948. 
Figure (\ref{figV}) shows these data. Unfortunately, the large 
(or completely undetermined as in the case 
of 1948 data point) uncertainties of these measurements do not allow 
us to infer the values of $\beta_p$ and $\beta_s$ from these data 
alone. However, even the weak constraints based on these 
measurements turn out being quite useful.

We will use the fact that based on these data 
$v_{rot} \sin \beta$ is currently {\it increasing} 
for both primary and secondary components. As $v_{rot}$
is constant this can only be due to $\sin \beta_j$
increasing in time. From the evolution equation 
(\ref{eq:beta_dot}) we find the 
expression for the derivative of $\sin \beta_{j}$ ($j=p,s$):
\begin{eqnarray}
\frac{d\sin\beta_j}{dt} = 3 \Omega_{P,j} \sin i 
\sin \lambda_{j} \cos \beta_{j} (\cos i\cos\beta_j+\nonumber\\
+ \sin i\sin\beta_j\cos\lambda_j) \approx \frac{3}{2} \Omega_{P,j} 
\sin^2 i \sin (2\lambda_{j}) \cos \beta_{j},
\end{eqnarray}
where $\Omega_{P,j}$ is the frequency describing spin 
precession caused by the rotation-induced 
stellar oblateness (see equation (\ref{eq:Omega_P})) and 
the approximation holds for $i\approx 90^\circ$. 
From the Rossiter-McLaughlin measurements of Albrecht \etal (2009)
we know that $\sin 2\lambda_j$ is positive for the primary 
and negative for the secondary. Thus, it follows from the current 
time derivatives of $\sin\beta_{p,s}$ that $\beta_{p}$ must be less 
than $90^\circ$ while $\beta_{s}$ should be greater than 
$90^\circ$. We use this information to help constrain stellar 
spin orientation in \S \ref{sect:results}.


\subsection{Photometric fitting procedure.}  
\label{sec:fit}

The specific parameters of DI Herculis that we used in simulations 
are summarized in Table \ref{tbl:pars}. To keep things simple we
have only varied the most important unknown quantities --- the angles
$\beta_p$ and $\beta_s$. Given the quality of the data we expect 
that fitting for extra parameters in our model would result in
too many degeneracies between the different model variables.

We integrate back in time the evolution equations for spin 
and orbital parameters (Appendix \ref{app:evolve}) starting 
from 15 July, 2008, which is set as the initial point in our 
calculations. The angles $\beta_p$ and $\beta_s$ that we vary 
correspond to this epoch. The other two angles specifying spin 
orientation were fixed at $\lambda_p=72^\circ$ and 
$\lambda_s=-84^\circ$ based on the Rossiter-McLaughlin measurements 
of Albrecht \etal (2011).

To obtain better eclipse fits we had to introduce quite different 
limb-darkening coefficients for the two components, which, given 
the proximity of stellar masses in DI Her, suggests that our data 
are affected by some systematic effects. Nevertheless, given that 
the limb darkening only weakly affect the non-axisymmetric surface
brightness distribution due to gravity darkening (see 
\S \ref{sect:int_dist}), the actual values of limb-darkening 
coefficients are not so important. 

We constrain DI Her spin orientation as follows. For each pair 
$\beta_p$, $\beta_s$ we compute
\begin{equation}
\chi^2 =\frac{1}{N}\sum\limits_{j=1}^9\sum\limits_{i=1}^{N_j} 
\frac{(I(t_i) - I^j_{obs}(t_i))^2}{\sigma^2_i},
\label{eq:chi}
\end{equation}
where index $j$ runs through all 9 eclipses, $i$ runs through 
the number of data points per each lightcurve (total of 
$N_j$ for $j$-th eclipse), $N=\sum_{j=1}^9 N_j$ is the total 
number of data points, $I(t)$ is given by equation 
(\ref{eq:intensity}), $I^j_{obs}(t)$ is the observed intensity, 
and $\sigma_i$ is the variance. The best
fit values of $\beta_p$ and $\beta_s$ are determined by finding the
minimum of $\chi^2$ over a large two-dimensional grid of values
of these angles. We use only the eclipses with well-defined minima.
We did not try to match theoretical and observational eclipse 
minima with our direct backward integration in time and instead 
just shift theoretical curves horizontally by small amount at 
each epoch for a better fit.

\begin{figure}
\epsscale{1.1}
\plotone{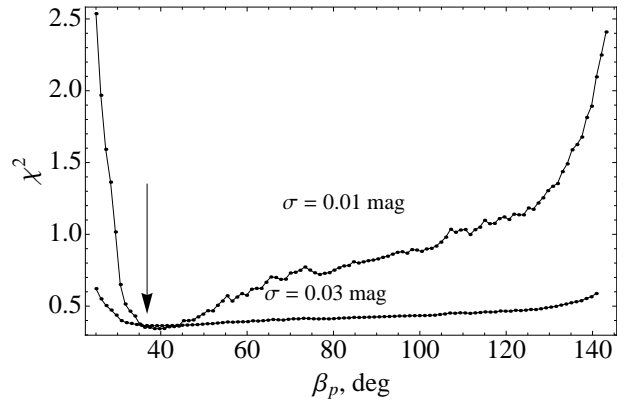}
\caption{Sensitivity of spin angle determination to the noise
level assumed for the data. We simulate the primary eclipse lightcurve 
using theoretical prescription (\ref{eq:intensity}) with a 
given level of the Gaussian noise $\sigma$ for a fixed spin angle 
$\beta_p=40^\circ$ for the primary. We vary only this angle
in our $\chi^2$ fitting ($\beta_s=40^\circ$, $\lambda_p=72^\circ$ and $\lambda_s=-84^\circ$ are constant here) just to illustrate that large noise level
does not allow us to constrain system parameters, while 
$\sigma\approx 0.01$ mag  yields the correct orientation of the system.}
\label{fig4}
\end{figure}

As mentioned before, the errorbars for our dataset are not 
constrained, so we employed the following procedure to estimate 
them. First, we took all $\sigma_i$ to be constant and run our
minimization procedure to find the best fit values of $\beta_p$
and $\beta_s$. Second, for each out of 9 eclipses we measure
the scatter $\sigma_j$ of the observational data points around 
the model lightcurve computed assuming these particular values 
of $\beta_p$ and $\beta_s$. This provides us with 9 different 
values of $\sigma_j$ (indicated in panels in Figure \ref{fig5} for 
each eclipse), which we use as error estimates in equation 
(\ref{eq:chi}). Typical values of $\sigma_j$ found using this 
procedure are $\sim 0.01$ mag. We then perform the final 
$\chi^2$ minimization adopting these values of $\sigma_j$
as error estimates for corresponding eclipses. In this approach 
all data points corresponding to $j$-th eclipse have a single 
value of the photometric error equal to $\sigma_j$. 

We test the performance of our fitting algorithm by applying
it to a simulated dataset in a simplified setup. We calculate 
a theoretical primary eclipse lightcurve including the gravity 
darkening effect and assuming a binary with physical parameters 
($M_\star,R_\star,T_\star$, etc.)
of the DI Her (in particular with $S\approx 0.04$ for both 
stars). We take somewhat arbitrarily $\beta_p=40^\circ$, 
$\beta_s=40^\circ$, $\lambda_p=72^\circ$ and 
$\lambda_s=-84^\circ$. We then add some random Gaussian noise 
with variance $\sigma$ to this simulated lightcurve. For this 
test we assume $\beta_s$, $\lambda_p$ and $\lambda_s$ to be known 
and try to measure the value of only $\beta_p$ using our procedure.
As a consequence, we need to perform only one-dimensional 
minimization over $\beta_p$. 

The results of this exercise are shown in Figure \ref{fig4},
where we show $\chi^2$ curves for two different values
of the noise variance $\sigma$: $0.01$ mag and $0.03$ mag.
One can see that for $\sigma=0.03$ mag our parameter estimation 
procedure cannot recover the adopted value of $\beta_p$ --- the
$\chi^2$ distribution has very extended flat bottom which
does not lead to a useful constraint on $\beta_p$. However, 
for $\sigma=0.01$ mag our procedure works reasonably well and the 
minimim of $\chi^2$ is close to the input value of $\beta_p=40^\circ$. 
Since the simulated lightcurve was computed for realistic 
physical parameters of the DI Her and the noise levels for 
individual eclipses in real data $\sigma_j$ are indeed 
$\sim 0.01$ mag we expect that
our parameter estimation for a real dataset should be able
to determine real $\beta_p$ and $\beta_s$ with reasonable 
accuracy.


\section{Results}
\label{sect:results}


\begin{deluxetable}{ccc}
\tablecaption{DI Herculis parameters}
 \tablehead{\colhead{Parameter} & 
\colhead{Primary} & 
\colhead{Secondary}} 
\startdata
Stellar radius ($R_{\odot}$) &2.68&2.48\\
Stellar mass ($M_{\odot}$) &5.15&4.52\\
Von Zeipel's parameter $\beta_g$ & 0.1&0.1\\
Effective temperature (K) & 17300 & 15400\\
$v_{rot}\sin\beta$  (${\rm km \cdot s^{-1}}$) &108 & 116 \\
$c_1$ & 0.35 & 0.64\\
$\lambda$ ($^\circ$) & 72 &-84\\
\hline
Derived parameters & & \\
\hline\\
$\beta$ ($^\circ$) & $62\pm 17$ & $90\leq\beta_s\leq110$\\
$\omega R$ (${\rm km \cdot s^{-1}}$) &112 &124\\
S &0.040 &0.046\\
\enddata
\label{tbl:pars}
\end{deluxetable}

\begin{figure}
\epsscale{1.1}
\plotone{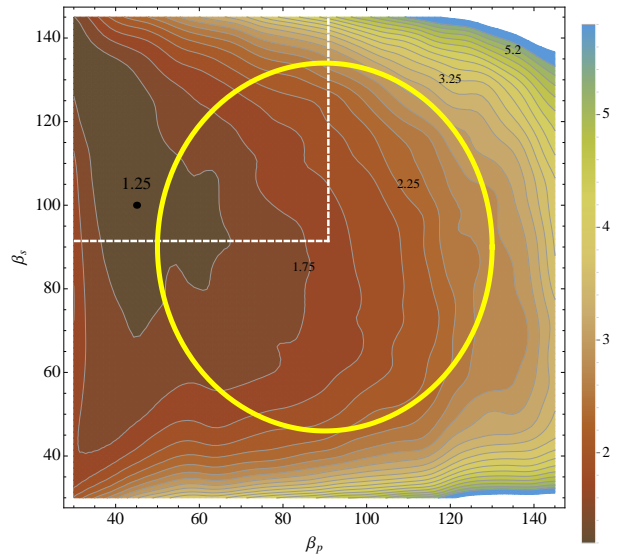}
\caption{$\chi^2$ distribution over all eclipses. The black dot 
shows the global minimum of $\chi^2$ distribution within the 
considered range of $\beta_p$ and $\beta_s$. The yellow 
ellipse shows the constrain coming from the precession rate 
corresponding to 1$\sigma$ level, where the analytical 
estimate of $\dot{\omega}$ is obtained using (\ref{eq:omega_dot}). 
Evolution of $v_{rot}\sin\beta$ at present time additionally 
constrains $\beta_p<90^\circ$,  $\beta_s>90^\circ$ 
(represented by white dashed lines). 
}
\label{fig6}
\end{figure}

We display the results of our fitting procedure in Figure \ref{fig6},
which shows a map of $\chi^2$ distribution as a function of 
$\beta_p$ and $\beta_s$. We see that in the broad region near the 
minimum the value of $\chi^2$ is almost constant, which precludes 
us from deriving accurate values of the spin angles from the eclipse 
analysis alone. While the angle $\beta_p$ for the primary is 
constrained to lie in the range $30^\circ -70^\circ$, the eclipse
photometry alone does not set a reasonable limit on $\beta_s$:
we can only say that it should lie within $50^\circ -140^\circ$ 
interval. This difference is caused by the different noise levels
for primary and secondary eclipses: 4 out of 5 secondary eclipses
used have $\sigma_j> 0.01$ mag, while 3 out of 4 primary eclipses 
have $\sigma_j<0.01$ mag (one primary eclipse has $\sigma_j=0.004$ 
mag). As we demonstrated in previous section, large values 
of $\sigma_j$ significantly deteriorate the performance of our
parameter estimation procedure (see Figure \ref{fig4}), which
is apparently the case for secondary eclipses, during which 
the lightcurve is most sensitive to $\beta_s$. 

To obtain a better measurement of these angles we apply two additional
constraints. One of them uses the observed precession rate 
$\dot{\omega}_{obs} = 1^{\circ}.24 \pm 0^{\circ}.18/100$ yr
(Martynov \& Khaliullin 1980). The apsidal precession rate $\dot{\omega}$ 
depends on $\beta_{p,s}$ since it contains a contribution 
due to the rotation-induced stellar quadrupole, see equation
(\ref{eq:omega_dot}), while the latter depends on these angles 
according to equation (\ref{eq:vsin}). We show the 
constraint on the apsidal precession rate (corresponding 
to 1$\sigma$ deviation) by yellow curve in Figure \ref{fig6}, where 
the analytical estimate of $\dot{\omega}$ is obtained using
(\ref{eq:omega_dot}).

\begin{figure}
\epsscale{1.15}
\plotone{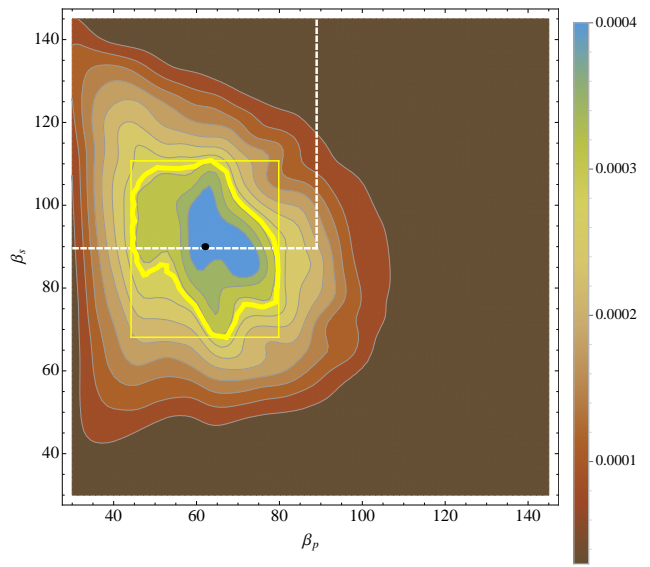}

\caption{Probability density distribution for $\beta_p$, $\beta_s$ obtained by combining the eclipse fitting and the constraint on the precession rate. Yellow thick contour gives $1\sigma$ level. The best fit values of spin  angles that we derive from this map are $\beta_p=62^\circ\pm 17^\circ$ and $\beta_s=90^\circ\pm 20^\circ$ (yellow thin contour). The white dashed curve additionally shows the $v_{rot}\sin\beta$ evolution constraint, see \S \ref{sect:vsini_evol}.}
\label{fig10}
\end{figure}

To obtain approximate values and the error bars of the angles 
$\beta_p$ and $\beta_s$ based on the eclipse fitting and the 
measurement $\dot \omega$, we first construct the photometric 
probability distribution of these angles using the $\chi^2$ map
from Figure \ref{fig6}. We then convolve it with the 
distribution of $\beta_p$ and $\beta_s$ (assumed to be a 
two-dimensional Gaussian) based on the $\dot \omega$
measurement of Albrecht \etal (2009). The map of the resultant 
probability density distribution is shown in Figure \ref{fig10}.
From this map we find $\beta_p=62^\circ\pm 17^\circ$ and 
$\beta_s=90^\circ\pm 20^\circ$, where the errors correspond to 
1-$\sigma$ uncertainty. Comparing with Figure \ref{fig6} we see 
that $\beta_s$ is constrained essentially purely by the $\dot\omega$ 
measurement, with photometric data not being useful. At the same time, 
for the primary angle $\beta_p$ the photometric data do result in 
a meaningful measurement, reducing $\beta_p$ from the value 
suggested by $\dot\omega$ alone and lowering error considerably.

Another constraint on spin orientation is based on the 
evolution of $v_{rot} \sin \beta$ for both components 
(see \S \ref{sect:vsini_evol}) and is illustrated by the 
white dashed line in Figures \ref{fig6} \& \ref{fig10}.  
This constraint is most important for the spin orientation of 
the secondary as it excludes $\beta_s < 90^\circ$ from the
 consideration. As a result, we come up with a refined measurement 
of $\beta_s=100^\circ\pm 10^\circ$. 


\begin{figure}
\epsscale{1.15}
\plotone{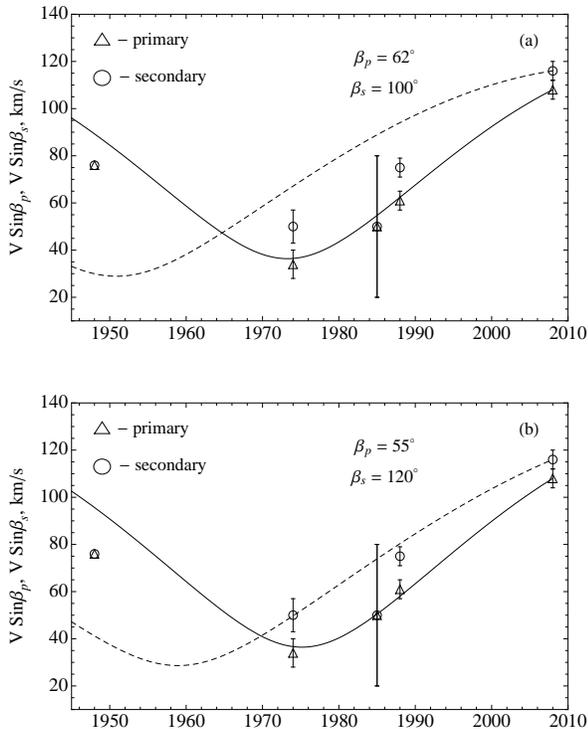}
\caption{Evolution of $V_{j} \sin \beta_{j}$ for both stars for derived angles $\beta_p=62^{\circ}$, $\beta_s=100^{\circ}$ (a) and for another set $\beta_p=55^{\circ}$, $\beta_s=120^{\circ}$ (b). 
In both cases $\lambda_p=70^{\circ}$ and $\lambda_s=-76^{\circ}$. Solid and dashed curves 
correspond to theoretical curves for primary and secondary stars.
Circles and triangles with errorbars (when available) represent 
the measurements for secondary and primary, correspondingly, 
taken from Albrecht \etal (2009). It shows that the derived $\beta_s$ from photometrical analysis does not provide the best fit for $V_{s} \sin \beta_{s}$ evolution.}
\label{figV}
\end{figure}

Figure \ref{fig5} shows model lightcurves for these best fit values
of $\beta_p$ and $\beta_s$ for all eclipses used in this work. One can 
clearly see the existence of some features in the lightcurves that 
remain unfit by our gravity-darkened model, especially at the 
midpoint of some eclipses. These are likely artefacts of the 
measurements using different instruments and at different 
locations.


\section{Discussion and conclusions}
\label{sect:discussion}


In this work we developed a method for determining full 
three-dimensional spin-orbit geometry of an eclipsing 
binary system with rapidly rotating components. The idea
behind this method lies in using the gravity darkening effect 
and its influence on the properties of the photometric 
lightcurve of the system. Using this method, coupled with 
two additional constraints --- the value of the apsidal 
precession rate of the system and the evolution of 
spectroscopically determined projection of the stellar 
rotation speed --- we were able to provide a reasonable 
measurement of the projections of stellar spins onto our 
line of sight in the eclipsing binary DI Her. 

A very similar technique based on the gravity darkening effect 
has already been employed to infer the spin-orbit orientation 
in a planetary system KOI-13.01, which contains a rapidly rotating 
($v_{rot}\sin\beta=65\pm 10$ km s$^{-1}$) intermediate mass 
star (Szab\'o \etal 2011; Barnes \etal 2011). This measurement 
used several eclipses (transits) obtained over a short time 
span, as opposed to our procedure that uses data spread over 
a long time interval. In the case of Barnes \etal (2011)
the exquisite photometric accuracy of {\it Kepler}
allowed derivation of a rather tight constraint on the 
spin orientation of the host star in KOI-13.01 system, 
something that we cannot accomplish with our low-quality 
multi-epoch photometry. The same kind of photometric
accuracy ($\sim 10^{-4}$ mag) would provide us with a much
better constraint on the DI Her orientation angles even with
single epoch data, see \S \ref{sect:int_dist}. 

Modeling the gravity darkening-modified eclipse lightcurves in
binary stars is not an easy task because each component of 
the binary covers large portion of the disk of another. This 
requires integration of the intensity distribution over a large 
fraction of the stellar surface, which naturally gives rise to 
degeneracies between different parameters of the system, 
making it difficult to determine stellar spin orientation. 
On the other hand, in the case of planetary transits planet
covers only a small fraction of the stellar surface so that
the eclipse lightcurve can be directly related to the
one-dimensional run of stellar surface temperature asymmetries. 
The latter can be much more easily modeled via the gravity 
darkening effect to infer the system orientation. 

As of now there is no good explanation for the strong 
spin-orbit misalignment of DI Her (Albrecht \etal 2010). 
It could be primordial, resulting from an interaction 
between the stars and the disk from which they formed, 
which would require some yet unknown mechanism to get 
accomplished. Alternatively, the system may contain a
third body in a wider orbit as suggested by Kozyreva 
\& Bagaev (2009) based on timing of eclipses over 
a long time span. If the orbit of that body is highly 
inclined with respect to the orbit of the inner two 
stars then the Lidov-Kozai mechanism (Lidov 1962; Kozai 1962) 
may be invoked to explain the spin-orbit misalignment 
in DI Her. This idea clearly requires further investigation, 
but we will mention that Lidov-Kozai cycles with tidal dissipation 
are often considered responsible (Fabrycky \& Tremaine 2007) for  
the spin-orbit misalignments inferred in many 
extrasolar planetary systems (Albrecht \etal 2012).

It is interesting that recently started BANANA project 
(Albrecht \etal 2011) focusing on the Rossiter-McLaughlin 
measurements in binaries containing rapidly spinning stars 
reported close spin-orbit alignment for the primary star 
in the NY Cep system, which is similar to DI Her in 
many respects. On the other hand, there are other eclipsing 
binary systems such as AS Camelopardalis, which exhibit 
anomalously slow apsidal precession, similar to DI Her 
(Pavlovski \etal 2011). If it is the spin-orbit 
misalignment that is causing the anomalous precession in 
AS Camelopardalis then the photometric method developed 
in this work and Barnes (2009) may be used to constrain the system 
orientation (although the measured projected rate in this 
system is not very high, $\approx 15$ km s$^{-1}$ for 
the primary component).

\acknowledgements

Authors are grateful to the referee, Jason Barnes for valuable 
suggestions that helped us improve the manuscript, Valentina 
Kozyreva for providing us with observational data, and Ed 
Turner for useful 
discussions. AAP thanks Princeton University for hospitality
during the time when part of this work was performed. Financial 
support of this research is provided by the Dinasty fellowship 
for AAP and by the Sloan Foundation and NASA via grant NNX08AH87G 
for RRR. 


\appendix


\section{Geometry of the sky--projected stellar disc}
\label{eq:conv}

In this section we will derive the relation between the 
coordinates of a given point on the stellar surface in the  
observer and symmetry frames. Normal $\boldsymbol{\xi}$ to 
the stellar surface at a point $(x,y,z)$ in the symmetry 
frame is ($|\boldsymbol{\xi}|\neq 1$)
\begin{equation}
\boldsymbol{\xi}=
\left(\frac{x}{{\eta}^2}, \frac{y}{{\eta}^2},z\right)
\end{equation}
Because of the rotation-induced distortion the sky-plane 
projected stellar shape is not circular. The last visible 
points on the stellar surface for the observer are given 
by the following equation
\begin{equation}
{\bf n}\cdot{\boldsymbol{\xi}} = 0
\end{equation}
where $\bf n$ is a unit vector towards the observer given
by equation (\ref{observ}). It results in the following equations:
\begin{eqnarray}
z=\frac{\tan{\beta}}{\eta^2}x,~~~~~~~~~~
\frac{x^2}{\eta^2}\left(1+\frac{\tan^2\beta}{\eta^2}\right)+\frac{y^2}{\eta^2}=R^2_{pol}
\label{critic2}
\end{eqnarray}
Equations (\ref{critic2}) give us the "critical line" of the 
last visible points on the stellar surface (where $\boldsymbol{\xi}$
lies in the sky plane) in the symmetry system. It defines the 
shape of the visible stellar disk. To obtain the coordinates 
in observer frame the corresponding coordinate transformation
should be made:
\begin{eqnarray}
x=\cos\beta x_{0}-\sin\beta Z,~~~~~~~~
z=\sin\beta x_{0}+\cos\beta Z,
\label{eq:xz}
\end{eqnarray}
where we defined
\begin{eqnarray}
x_{0}=\cos\lambda X+\sin\lambda Y,~~~~~~~~
y_{0}=-\sin\lambda X+\cos\lambda Y.
\label{eq:xy0}
\end{eqnarray}
So the equation for the critical line written in the observer 
frame is
\begin{eqnarray}
Z=\frac{x_0(\eta^{-2}-1)\tan \beta }{1+\eta^{-2}\tan^2 \beta} \equiv x_{0}\tan p, \label{eq:Z1}\\
\frac{x^2_{0}}{\cos^2\beta +\eta^{-2}\sin^2\beta}+y^2_{0}=\eta^2 R^2_{pol}.
\end{eqnarray}
where $p$ is defined by equation (\ref{eq:Z1}). In coordinates $(x_0,y_0,Z)$ the stellar surface is described by
\begin{equation}
\frac{y^2_{0}+(\cos(\beta)x_{0}-\sin(\beta)Z)^2}{\eta^2}+\left(\sin(\beta)x_{0}+\cos(\beta)Z\right)^2=R^2_{pol},
\end{equation}
from which one can find $Z$ in terms of $X$ and $Y$. The solution 
to the resulting quadratic is
\begin{equation}
Z=x_{0}\tan p+\frac{\eta^{-1}\sqrt{det}}
{\cos^2\beta+\eta^{-2}\sin^2\beta}
\label{eq:Z}
\end{equation}
where the expression for the determinant $det$ is
\begin{equation}
{det}=-\left[x^2_{0}+\left(y^2_{0}-\eta^2 R^2_{pol}\right)\left(\cos^2\beta+\frac{\sin^2\beta}{\eta^2}\right)\right].
\label{eq:det}
\end{equation}
It can be easily checked that the condition $det=0$ coincides 
with the equation for the critical line, so in the region 
interior to this line $det>0$. We choose the positive root of 
the determinant (the negative root represents the invisible 
side of the star as seen from Earth, see Barnes (2009). So if 
$(X,Y,0)$ is the point on the sky-projected stellar disc then 
${\bf R}(X,Y)=(X,Y,Z(X,Y))$ represents the full three-dimensional
coordinates of the point {\it on the stellar surface} right above
$(X,Y,0)$ in the observer frame.

In general, to find coordinates $(x,y,z)$ of a point on the 
stellar surface in the symmetry frame 
corresponding to a point $(X,Y)$ projected onto the sky plane 
one has to compute $x_0$ and $y_0$ via equation (\ref{eq:xy0}),
determine $Z$ using (\ref{eq:Z})-(\ref{eq:det}), obtain $x$ and 
$z$ from equation (\ref{eq:xz}), and finally determine $y$
from equation (\ref{form}).


\section{Time evolution} 
\label{app:evolve}

Here we derive equations that describe the evolution of the binary 
orbit and spin orientation of its components. We denote 
${\bf L}=L{\bf l}$, ${\bf S}_j=S_j{\bf s}_j$, $j=p,s$
the orbital angular momentum and spin angular momenta of the two 
stars respectively, $|{\bf l}|=|{\bf s}_j|=1$. Here 
\ba
L=\mu\Omega_K a^2\sqrt{1-e^2},~~~S_j=I_j\omega_j=\eta_j M_j R_j^2\omega_j,
\label{eq:momenta}
\ea
where $M_j$, $R_j$, $\omega_j$ are the masses, radii and spin 
angular frequencies  of the two stars, $\eta_j$ are constants 
determining their moments of inertia, $\mu=M_p M_s/(M_p+M_s)$ is
the reduced mass, $a$, $e$, and $\Omega_K=[G(M_p+M_s)/a^3]^{1/2}$
are the semi-major axis, eccentricity and mean orbital frequency.
We introduce unit vector ${\bf n}$ from the system's barycenter 
to the observer and direct a Cartesian coordinate system in the 
directions ${\bf n}$, ${\bf m}$, ${\bf k}$
where
\ba
{\bf k}=\frac{{\bf n}\times {\bf l}}{\sin i},~~~
{\bf m}={\bf k}\times {\bf n}=\frac{1}{\sin i}[{\bf l}-
{\bf n}({\bf l}\cdot {\bf n})],
\label{eq:orts}
\ea
where $i$ is the inclination of the system, $\cos i=({\bf l}\cdot 
{\bf n})$, see Figure \ref{fig1}.

Orientation of ${\bf s}_j$ is conventionally given by the three
angles $\alpha_j$, $\beta_j$, $\gamma_j$ between each of  ${\bf s}_j$
and vectors ${\bf l}$, ${\bf n}$, and ${\bf k}$ correspondingly,
i.e.
\ba
\cos\alpha_j=({\bf s}_j\cdot {\bf l}),~~~
\cos\beta_j=({\bf s}_j\cdot {\bf n}),~~~
\cos\gamma_j=({\bf s}_j\cdot {\bf k}).
\label{eq:angle_defs}
\ea
Then it is easy to show that
\ba
{\bf s}_j=\cos\beta_j{\bf n}+\frac{\cos\alpha_j-\cos i\cos\beta_j}
{\sin i}{\bf m}+\cos\gamma_j{\bf k}.
\label{eq:s_expr}
\ea

Observationally, it is also convenient to introduce angle $\lambda_j$
between the projection ${\bf s}_{\perp,j}=[{\bf s}_j-
{\bf n}({\bf s}_j\cdot {\bf n})]/\sin\beta_j$ of ${\bf s}_j$ onto the 
sky plane and the projection ${\bf m}$ of vector ${\bf l}$ onto the 
same plane: $\cos\lambda_j=({\bf s}_{\perp,j}\cdot {\bf m})$. 
This is the angle which is measured by the Rossiter-McLaughlin 
effect. One can easily show that these four angles are related via
\ba
\cos\alpha_j=\cos i\cos\beta_j+\sin i\sin\beta_j\cos\lambda_j,~~~~~
\cos\gamma=\sin\lambda\sin\beta.
\ea
Thus, knowing $\lambda_j$, $\beta_j$ and $i$ one can immediately
obtain $\alpha_j$ from these expressions.

Orientation of the orbital ellipse in the plane of the orbit is 
given by the eccentricity vector ${\bf E}=e{\bf e}$ which points 
from the main focus to the pericenter. Given that direction defined 
by vector ${\bf k}$ is the direction of the line of nodes we identify the
angle between ${\bf k}$ and ${\bf e}$ as the longitude of the periastron 
$\omega$ and write
\ba
{\bf e}=\cos\omega{\bf k}+\sin\omega ({\bf l}\times {\bf k}).
\label{eq:e_def}
\ea

Stellar asphericity due to rotation and tides as well as relativistic 
effects lead to evolution of ${\bf l}$, ${\bf e}$, and ${\bf s}_j$
described by the following equations (Barker \& O'Connell 1975):
\ba
\dot{\bf l}={\bf \Omega}_L\times {\bf l},~~~~~~
\dot{\bf e}={\bf \Omega}_L\times {\bf e},~~~~~~
\dot{\bf s}_j={\bf \Omega}_{S,j}\times {\bf s}_j,
\label{eq:prec_s}
\ea
where
\ba
&& {\bf \Omega}_L=\left(\Omega_E+\sum\limits_{j=s,p}\Omega_{T,j}\right)
{\bf l}+\sum\limits_{j=s,p}\Omega_{Q,s}
\left[\cos\alpha_j {\bf s}_j+\frac{1-5\cos^2\alpha_j}{2}{\bf l}\right],
\label{eq:Omega_L}\\
&& {\bf \Omega}_{S,j}=\Omega_{G,j}{\bf l}+\Omega_{P,j}
\left({\bf s}_j-3\cos\alpha_j{\bf l}\right)\approx\Omega_{P,j}
\left({\bf s}_j-3\cos\alpha_j{\bf l}\right).
\label{eq:Omega_S}
\ea
Here different contributions to precession rates are denoted
as follows (Barker \& O'Connell 1975; Claret et al 2010):
orbital Einstein precession
\ba
\Omega_E=\dot\omega_{\rm GR}=\frac{3G\Omega_K(M_s+M_p)}{c^2 a(1-e^2)}\approx 
2.35^\circ/100~\mbox{yr},
\label{eq:Omega_E}
\ea
orbital precession caused by stellar quadrupole due to
tidal distortions ($k_{2,j}$ are introduced below)
\ba
&& \Omega_{T,j}=15k_{2,j}\Omega_K\frac{M_r}{M_j}\left(\frac{R_j}{a}\right)^5
\frac{8+12e^2+e^4}{8(1-e^2)},~~~r\neq j,
\label{eq:Omega_T}
\\
&&\Omega_{T,p}\approx 
0.69^\circ/100~\mbox{yr},
~~~\Omega_{T,s}\approx 
0.63^\circ/100~\mbox{yr},
\nonumber
\ea
orbital precession due to rotation-induced stellar quadrupole 
\ba
&&\Omega_{Q,j}=-\frac{3}{2}\frac{G(M_p+M_s)J_{2,j}}{\Omega_K a^5(1-e^2)^2}=
-k_{2,j}\frac{M_p+M_s}{M_j}\frac{\omega_j^2}{\Omega_K(1-e^2)^2}
\left(\frac{R_j}{a}\right)^5\label{eq:Omega_Q}
\\
&&\Omega_{Q,p}\approx 
-\frac{2.1^\circ/100~\mbox{yr}}{(\sin\beta_p)^2}s^{-1},
~~~\Omega_{Q,s}\approx 
-\frac{2.2^\circ/100~\mbox{yr}}{(\sin\beta_s)^2}s^{-1},
\nonumber
\ea
geodetic spin precession
\ba
&&\Omega_{G,j}=\frac{G\Omega_K\mu(4+3M_r/M_j)}{2c^2 a(1-e^2)},~~~j\neq r,
\label{eq:Omega_G}
\\
&&\Omega_{G,p}\approx 
0.61^\circ/100~\mbox{yr},~~~
\Omega_{G,s}\approx 
0.69^\circ/100~\mbox{yr},
\nonumber
\ea
and the spin precession caused by rotation-induced stellar oblateness 
\ba
&&\Omega_{P,j}=\frac{GM_pM_sJ_{2,j}}{2I_j\omega_j a^3(1-e^2)^{3/2}}=
\frac{k_{2,j}}{3\eta_j}\frac{M_r}{M_j}\frac{\omega_j}{(1-e^2)^{3/2}}
\left(\frac{R_j}{a}\right)^3,~~~r\neq j,
\label{eq:Omega_P}
\\
&&\Omega_{P,p}\approx
\frac{136.6^\circ/100~\mbox{yr}}{\sin\beta_p},~~~
\Omega_{P,s}\approx
\frac{168.6^\circ/100~\mbox{yr}}{\sin\beta_s}s^{-1},
\nonumber
\ea
In equations (\ref{eq:Omega_T}), (\ref{eq:Omega_Q}), (\ref{eq:Omega_P})
$k_{2,j}$ is the apsidal motion constant related to stellar 
rotation-induced quadrupole moment constant $J_{2,j}$ via
\ba
k_{2,j}=\frac{3}{2}\frac{J_{2,j}}{R_j^2}
\left(\frac{\omega_{b,j}}{\omega_j}\right)^2,~~~
\omega_{b,j}\equiv\left(\frac{GM_j}{R_j^3}\right)^{1/2},
\ea
with $\omega_{b,j}$ being the breakup angular frequency. Numerical
estimates assume $k_{2,j}\approx 0.008$ (Claret \etal 2010) 
and the moment of inertia constant $\eta_j=0.063$ 
(Claret \& Gimenez 1989).
Given that $\Omega_{G,j}\ll \Omega_{P,j}$ we dropped geodetic 
contribution in equation (\ref{eq:Omega_S}). 

Differentiating relation $\cos i=({\bf l}\cdot 
{\bf n})$ with respect to time and using equations 
(\ref{eq:orts}), (\ref{eq:angle_defs}), (\ref{eq:prec_s}), 
and (\ref{eq:Omega_L}) one obtains 
\ba
\dot i=-\frac{\dot{\bf l}\cdot {\bf n}}{\sin i}={\bf \Omega}_L
\cdot {\bf k}=\sum\limits_{j=p,s}\Omega_{Q,j}
\cos\alpha_j\cos\gamma_j.
\label{eq:i_dot}
\ea
Because of precession of ${\bf l}$ vectors ${\bf k}$
and ${\bf m}$ vary in time. Differentiating equations (\ref{eq:orts})
with respect to time and using (\ref{eq:angle_defs}), (\ref{eq:prec_s}), 
and (\ref{eq:Omega_L}) their evolution is governed by 
equations
\ba
\dot {\bf k}=\frac{{\bf m}}{\sin^2 i}\sum\limits_{j=p,s}\Omega_{Q,j}
\cos\alpha_j\left(\cos i\cos\alpha_j-\cos\beta_j\right),
~~~~~~~\dot {\bf m}=\frac{\bf k}{\sin^2 i}\sum\limits_{j=p,s}\Omega_{Q,j}
\cos\alpha_j\left(\cos\beta_j-\cos i\cos\alpha_j\right),\label{eq:dotm}
\ea

Next, we differentiate $\cos\omega=({\bf e}\cdot{\bf k})$ as well as 
each of the relations (\ref{eq:angle_defs}) with respect to time
and transform them using equations (\ref{eq:orts})-(\ref{eq:s_expr}), 
(\ref{eq:e_def})-(\ref{eq:Omega_S}), (\ref{eq:i_dot}) and 
(\ref{eq:dotm}). As a result we arrive at the following 
expressions: 
\ba
\dot\omega &=& -\frac{(\dot{\bf e}\cdot {\bf k})+({\bf e}\cdot \dot{\bf k})}
{\sin\omega}
\nonumber
\\
&=& 
=\Omega_E+\sum\limits_{j=p,s}\left\{\Omega_{T,j}+
\frac{\Omega_{Q,j}}{\sin^2 i}\left[
\cos\alpha_j(\cos\alpha_j-\cos i\cos\beta_j)+\sin^2 i
\frac{1-5\cos^2\alpha_j}{2}\right]\right\},\label{eq:omega_dot}
\\
\dot\alpha_j &=& -\frac{(\dot{\bf l}\cdot {\bf s}_j)+
({\bf l}\cdot \dot{\bf s}_j)}
{\sin\alpha_j}
\nonumber
\\
&=&-\Omega_{Q,r}\frac{\cos\alpha_r}{\sin i\sin\alpha_j}\left[
\cos\gamma_r(\cos i\cos\alpha_j-\cos\beta_j)-\cos\gamma_j
(\cos i\cos\alpha_r-\cos\beta_r)\right],~~~j\neq r,\label{eq:alpha_dot}
\\
\dot\beta_j &=& -\frac{(\dot{\bf s}_j\cdot {\bf n})}
{\sin\beta_j}
=3\Omega_{P,j}\frac{\sin i\cos\gamma_j\cos\alpha_j}
{\sin\beta_j},\label{eq:beta_dot}
\\
\dot\gamma_j &=& -\frac{(\dot{\bf k}\cdot {\bf s}_i)+
({\bf k}\cdot \dot{\bf s}_i)}
{\sin\gamma_i}\nonumber
\\
&=&-\frac{\cos\alpha_j-\cos i\cos\beta_j}{\sin^3 i\sin\gamma_j}
\sum\limits_{r=s,p}\Omega_{Q,r}\cos\alpha_r
(\cos i\cos\alpha_r-\cos\beta_r)
\nonumber\\
&&
+\frac{3\Omega_{P,j}\cos\alpha_j
(\cos i\cos\alpha_j-\cos\beta_j)}{\sin i\sin\gamma_j}.
\label{eq:gamma_dot}
\ea
Equations (\ref{eq:i_dot}), (\ref{eq:omega_dot})-(\ref{eq:gamma_dot})
constitute a closed system of 8 evolution equations for 8 unknown
angles --- $i$, $\omega$, $\alpha_j$, $\beta_j$, $\gamma_j$, $j=p,s$ 
--- fully determining the orbital orientation of the binary and  
spin orientation of each star.

One can check the validity of these expressions by using the fact that
the total angular momentum of the system ${\bf J}=L{\bf l}+S_p{\bf s}_p
+S_s{\bf s}_s$ is conserved. Equation (\ref{eq:omega_dot}) agrees 
with the analogous expression in Shakura (1985).



\end{document}